**$MoS_2$/C @ Polyurethane Composite Sponges for Synergistic High-Rate Solar Steaming and Mercury Removal**

*Weigu Li,*[1] *Marshall C Tekell,*[2] *Yun Huang,*[3] *Karina Bertelsmann,*[1] *Max Lau,*[1] *and Donglei Fan*[1,3,*]

[1]Department of Mechanical Engineering, The University of Texas at Austin, Austin, TX 78712, USA

[2]Department of Chemical Engineering, The University of Texas at Austin, Austin, TX 78712, USA

[3]Materials Science and Engineering Program, Texas Materials Institute, The University of Texas at Austin, Austin, TX 78712, USA

**Keyword**

Solar steaming, water treatment, $MoS_2$, Mercury, pathogen removal

## Abstract

Solar steam generation, a sustainable water-purification technology, holds substantial promises in resolving the global issue of shortage of drinkable water. Here, we report the design, fabrication, and performance of an innovative three-dimensional (3-D) solar steamer, offering synergistic high-rate steaming and heavy metal removal functions. The device is made of synthesized carbon-molybdenum-disulfide microbeads electrostatically assembled on a 3-D polyurethane sponge. The mesoporous composite sponge also serves as a freestanding water reservoir that avoids one-side contact to bulk water, effectively suppressing the commonly observed parasitic heat loss, and offering a high energy efficiency of 88%. When being sculptured into a 3-D spoke-like structure, the composite sponge achieves one of the highest evaporation rates of 1.95 kg $m^{-2}$ $h^{-1}$ at 1 sun. The solar steamer is demonstrated for water treatment, *i.e.* decontamination of metal ions, disinfection, and reducing alkalinity and hardness of river water. Particularly, the strong mercury adsorption of $MoS_2$ reduces Mercury from 200 to 1ppb, meeting the stringent standard set by the Environmental Protection Agency, which is the first demonstration of mercury-removal powered by solar energy. The unique design, fabrication, water-handling strategy, and mercury-removal function of this high-performance solar steamer could inspire new paradigms of water treatment technologies.

## 1. Introduction

The scarcity of potable water is an acute and global issue. According to the United Nations, more than 800 million people have no access to safe drinking water. Waterborne diseases lead to over 10,000 deaths per day including 5,000 children under the age of five.[1] It is highly desirable to develop low-cost, robust, and efficient technologies to disinfect and decontaminate water at the point of use, which is especially urgent for developing countries, remote areas without basic infrastructures, and people who cannot evacuate during natural disasters.[2-5] Solar steam generation that converts solar energy to heat and thereby produces clean water with a minimal carbon footprint[6] has emerged as one of the most promising sustainable technologies for water purification, sterilization, and desalination.[7-12] In the last five years, great advances have been made on the development of photothermal materials, ranging from plasmonic substrates to carbonaceous materials, such as metallic nanoparticles,[7, 10, 13, 14] metal-salts/oxides,[15, 16] graphene/graphite,[11, 17-19] graphene oxide (GO),[6, 20-25] and carbon nanotubes (CNTs).[21, 22, 26, 27] The efficiency of solar steaming is governed by four factors: efficiency and bandwidth of solar absorption, heat localization, hydrophilicity for water transportation, and paths of water evaporation.[6, 11, 15, 17, 25, 27-30] Recently, great attention has been focused on the optimization of thermal management of sunlight absorbers, the structural design of which evolved from suspending nanoparticles,[9, 31-33] to composite foams/films, to heat-loss-suppressing bilayer structures. [14, 17, 18, 27] In most studies of the bilayer structures, however, the two-dimensional (2-D) water transpiration path adversely curtails the available steam evaporation paths, limiting the water evaporation rate.[25, 29]

Contamination of water with heavy metal ions (e.g., Hg(II), Cr(VI), As(III), etc.) in developing and industrialized nations poses a serious environmental and health threat.[34] Particularly, mercury can permanently impair brain function in both children and adults, known as Minamata disease, via direct contact or bioaccumulation.[34-36] It is extremely important to remove mercury from aqueous resources effectively and facilely. Amidst a variety of water treatment technologies, adsorption emerges as one of the most effective methods to remove toxic mercury ions.[34, 37] Owing to the strong soft-soft interactions between mercury and sulfur,[38] sulfur-containing materials demonstrate outstanding performances in mercury removal.[37, 39-41] In particular, $MoS_2$ has proven its effectiveness in the purification of mercury polluted wastewater and drinking water, owing to its abundant sulfur atoms with their high mercury affinity.[37, 42-44] However, the necessity in removal of the mercury-enriched adsorbent composites can greatly complicate the process.[34]

In this work, we report an innovative solar steaming system, designed with consideration in providing high water evaporation rates, synergistic mercury removal functions, and manufacturability. The devices are strategically fabricated by assembling hydrothermally synthesized $MoS_2$/C microbeads, the dual-functional light-absorbing and mercury-removal material, on 3-D PU sponges ($MoS_2$/C @ PU). The synthesis of $MoS_2$ nanosheets on amorphous carbon microbeads renders the desired surface hydrophilicity for excellent water wettability, and modified zeta potential for tight attachment to the PU sponges by electrostatic assembling. The obtained $MoS_2$/C @ PU sponge offers a solar absorption efficiency of 98%. The 3-D evaporation path of a sculptured PU sponge effectively improves the evaporation rate to at least 1.95 kg $m^{-2}$ $h^{-1}$ under one sun, outperforming most reported works. The 87% of porosity and tunable thickness of the $MoS_2$/C @ PU sponge also retain water for storage and long-distance transportation, through which the heat loss to bulk water can be suppressed, enabling an energy conversion efficiency as high as 88%. When applied to treat natural river water, the solar steamer sterilizes bacteria, alleviates alkalinity, and reduces hardness. A single steaming operation also reduces 99.6% of metal ions ($Ni^{2+}$). Particularly, the rich sulfur atoms in $MoS_2$ effectively adsorb mercury from water during the solar-steaming process. Without residual absorbers, the harmful mercury is reduced from 200 ppb to the drinkable level of 1 ppb, meeting the standard set by the Environmental Protection Agency (EPA), which is the first demonstration of solar-powered mercury removal system.

## 2. Results and Discussions

### 2.1. Characterization of $MoS_2$/C Microbeads and MoS2/C @ PU composite Sponge

The fabrication of $MoS_2$/C @ PU composite starts with the hydrothermal growth of $MoS_2$/C microbeads followed by electrostatic assembling on a polyelectrolyte functionalized PU sponge, which is detailed in the experimental section. Carbonaceous materials have been extensively investigated owing to their broad applications in energy and environment.[11, 17, 18, 31, 45-49] Particularly, carbon beads or particles that provide high photo-thermal conversion efficiency in the near infrared region (NIR),[31, 48] and can be synthesized facilely and at a low cost, have received substantial research interest for solar-thermal applications. Here, we synthesized pure carbon microbeads and those with integrated vertical $MoS_2$ nanosheets ($MoS_2$/C) via a modified one-step hydrothermal reaction.[50] The roughened surfaces of $MoS_2$/C microspheres **[Figure 1a**] compared to that of pure carbon microbeads enhance the specific areas and solar absorption efficiency [Figure S1a-b].[11] The Raman spectrum confirms the chemical composition of the microbeads made of $MoS_2$/C **[Figure 1b]**. The two characteristic peaks at 378 and 404 $cm^{-1}$ correspond to the in-plane ($E_{2g}^1$) and vertical plane ($A_{1g}$) vibrational modes of Mo-S bonds in 2H-$MoS_2$, respectively.[51] Another two broad peaks at ~1357 and ~1589 $cm^{-1}$ correspond to the D and G bands of carbons,[52] respectively, which are also observed in the pure carbon microbeads [Figure S1c]. The crystalline structures of $MoS_2$/C microbeads are characterized by X-ray diffraction (XRD) spectroscopy as shown in Figure S1d. The characteristic peaks of $MoS_2$ nanosheets observed at 32.74° and 57.46° correspond to (100) and (110) planes, respectively, agreeing with previous reports.[53, 54] The absence of XRD diffraction peaks of crystalline carbon, *e.g.* at ~25° or ~ 44°, suggests the obtained carbon microbeads are largely amorphous.[55] The X-ray photoelectron spectroscopy (XPS) corroborates the chemical state of Mo and S. The two peaks with binding energies of 232.6 eV and 229.6 eV [Figure S2a] can be attributed to the doublet of Mo $3d_{3/2}$ and Mo $3d_{5/2}$, respectively.[51, 56] The other two peaks at about 162.6 eV and 161.5 eV [Figure S2b] arise from S $2p_{1/2}$ and S $2p_{3/2}$, respectively.[51, 56] The above characterizations confirm the unique chemistry and structure of the synthesized $MoS_2$/C microbeads as shown in **Figure 1c**.

Polyurethane sponge (PU) is chosen as the 3-D scaffold of the solar steaming system, not only because of the porous structures that provide large surface areas and numerous ramified microchannels for water storage, treatment, and evaporation, but also due to its extremely low thermal conductivity (<0.03 W $m^{-1}$ $K^{-1}$)[57, 58], which is less than those of GO films (~0.2 W $m^{-1}$ $K^{-1}$),[6] CNT composites (~0.2 W $m^{-1}$ $K^{-1}$),[27] and graphene foams (~1 W $m^{-1}$ $K^{-1}$).[18] The cost is also ultralow, often considered negligible as it is frequently used for packaging. However, the surface of PU is hydrophobic, and therefore resists water.

Here, we strategically synthesized amorphous carbon microbeads with integrated $MoS_2$. The presence of $MoS_2$ can not only effectively remove toxic mercury ions, but also alter the overall zeta potential of microbeads from 0 mV to a negative value of -26.6 mV [**Figure 1d**], which is critical for the successful assembling of $MoS_2$/C on the PU sponge. Since the surface of $MoS_2$/C microbeads is negatively charged, we functionalized the PU sponge with positive charges by polyelectrolytes via the layer-by-layer process as shown in **Figure 1e** (details in experimental section). Owing to the electrostatic interactions, the $MoS_2$/C microbeads tightly anchor onto the entire 3-D PU surfaces compactly and uniformly [**Figure 1f-h**]. The demonstrated electrostatic assembling of solar-thermal nanoparticles for solar steamers is a general approach, which could be applicable to other substrates, such as polymers,[12] woods,[10] and graphene/graphite,[59] for optimized applications.

Here, as show in **Figure 1g-h**, the microbeads maintain the roughened surfaces, which guarantee effective photo-thermal interaction areas and abundant active sites for removal of mercury. The resulting 3-D $MoS_2$/C@ PU composite exhibits excellent wettability, which instantly absorbs water droplets as shown in the slow-motion movie [Movie S1]. This high hydrophilicity is essential for an intimate contact of water with the heat-absorbing surfaces and its swift transpiration during steaming. In contrast, water droplets cannot penetrate pure PU sponges at all as shown in Movie S2.

The assembling of $MoS_2$/C microbeads changes the color of PU sponges into ultra-black [**Figure 1e**], which corroborates with the UV-visible-NIR measurement from 200 to 2500 nm. As shown in **Figure 1i**, with $MoS_2$/C microbeads, the average light absorption of PU can be substantially increased from ~47-58% to 98% in the spectrum range of 500 nm to 2 μm [**Figure 1i**]. Here, the high solar absorption due to $MoS_2$/C shows a difference of less than 4% in optical spectra when the thickness of a sponge is increased from 3 mm to 10 mm [Figure S3].

## 2.2. Solar Steaming Performances

2.2.1. Optimize Designed for High Evaporation Rates

Thermal management and water transport are critical factors in determining the overall performances of solar steaming. Here, five different designs are tested to manage heat dissipation and improve the evaporation pathway of water for large solar-thermal conversion efficiency and high water evaporation rate [**Figure 2a-(1-5)**]. The first setup in **Figure 2a-1** is similar to most state-of-the-art solar steamers, the thermally active material directly floats on bulk water.[11, 21] The second setup has a heat insulating layer (polystyrene foam) to reduce the heat dissipation from the thermally active 3-D sponge to the bulk water [**Figure 2a-2**];[6] a hydrophilic cellulose paper is used to wrap the insulating layer to assist water transport. This is named as indirect-contact setup with 2-D evaporation path. The third design is the same as the second one, except the 3-D sponge is exposed to air on both the top and sides, which we name as indirect-contact setup with 3-D evaporation path [**Figure 2a-3**]. In the fourth and fifth setups, the highly porous $MoS_2$/PU is directly employed as a freestanding water reservoir as well as a sunlight absorber, which is named as non-contact design [**Figure 2a-(4, 5)**]. A $MoS_2$/C @ PU sponge of 10 mm in thickness and 25 mm in diameter is tested to study the solar steaming performances of the above five designs. The typical results of time-dependent mass change under 1 sun (1 kW $m^{-2}$) are depicted in **Figure 2b**. The solar-assisted water evaporation rates are determined from the slope of the curves as shown in **Figure 2c**.[21] The intrinsic evaporation rate of pure water under one sun is 0.47 kg $m^{-2}$ $h^{-1}$. The utilization of $MoS_2$/C @ PU directly on bulk water [setup 1, **Figure 2a-1]** promotes the rate to 0.88 kg $m^{-2}$ $h^{-1}$. The utilization of a thermal insulating layer with hydrophilic paper for water transport [setup 2: indirect-contact with 2-D evaporation path, **Figure 2a-2**] increases the evaporation rate to 0.99 kg $m^{-2}$ $h^{-1}$. By exposing the entire 3-D light absorber sponge to air [setup 3: indirect-contact with 3-D evaporation path, **Figure 2a-3**], the evaporation rate is substantially improved by ~ 60% to 1.58 kg $m^{-2}$ $h^{-1}$, which can be mainly attributed to the additional evaporation paths via the sidewalls. Noticeably, when the 10 mm $MoS_2$/C @ PU absorber is directly applied as a water reservoir [setup 5: non-contact with 3-D evaporation path, **Figure 2a-5**], the evaporation rate is further elevated to 1.68 kg $m^{-2}$ $h^{-1}$, which could be attributed to the better thermal management in the reduction thermal loss at the bottom contact, as well as the much greater utilization of hot surfaces that directly transfer heat to water in the small pores compared to those in other designs. The same behaviors are confirmed in $MoS_2$/C @ PU sponges with a different thickness of 6 mm as shown in Figure S4.

The above results, determined from the time dependent mass change, are consistent with the temperatures obtained with infrared (IR) imaging. Here, we focus on comparing the two designs of setup 3 and 5 with the best performances. The thermal images of the 10 mm $MoS_2$/C @ PU in setup 3 **[Figure 2a-3**] and setup 5 [**Figure 2a-5**] are provided in **Figure 2d**. The highest temperature reached in the non-contact design [setup 5, **Figure 2a-5**], where the sponges are also used for water storage, is 38.5℃, ~ 1.1℃ higher than that obtained from the design in setup 3 [**Figure 2a-3**]. Along the cross-section, setup 5 [**Figure 2a-5**] shows a much more uniform temperature as observed in **Figure 2d-2** with an average value greater by 1.7℃, than that of the indirect-contact design in setup 3 shown in **Figure 2d-1**. It evidences the better controlled heat loss to the surrounding media, particularly at the bottom contact, in the design of setup-5, consistent with its overall greater performance compared to those of the other designs. These characterizations suggest that directly holding water in a solar steamer made of a low-mass sponge-like material, where the heat capacity is generally low, could greatly reduce the overall parasitic heat dissipation for high efficiency solar steaming.

As shown in the above, the highest performance is obtained from **setup 5**, the non-contact design with 3-D evaporation path [**Figure 2a-5**]. Further optimization of water evaporation paths by tuning the geometry of the photo-absorber could further improve water evaporation rate. Under 1 sun illumination, we tested this hypothesis by monotonically tuning the aspect ratio of $MoS_2$/C @ PU sponges via varying the thicknesses from 3 mm to 19 mm in setup 5 (non-contact design) and setup 3 (indirect-contact with 3-D evaporation path) and measured the time-dependent mass change as shown in **Figure 3a** and Figure S5a, respectively. The evaporation rates are calculated as shown in **Figure 3b**. In both designs, the increase in the thickness of $MoS_2$/C @ PU sponges always leads to notable improvements in evaporation rates. The 19 mm $MoS_2$/C @ PU in setup 5 (non-contact design), where water is absorbed in the pores of the sponge, offers the highest evaporation rate of 1.78 kg $m^{-2}$ $h^{-1}$, which is superior compared to most solar steaming materials reported in literature [**Figure 3c** and Table S1].

2.2.2. Analysis of the 3-D Evaporation Path

To quantitatively understand the geometrical factor of the 3-D evaporation solar-steaming sponge observed as above, we carried out the same time-dependent test in the dark (without solar illumination) as shown in **Figure 3d**. The analysis in **Figure 3e** determines a positive linear dependence between the evaporation rate in a dark condition ($r_{dark}$) on the aspect ratio of a sponge (t/D), following the relationship of $r_{dark} = 0.15 + \frac{1}{2} \cdot \frac{t}{D}$ (kg m$^{-2}$ h$^{-1}$), where $D$ and $t$ are the diameter and thickness of a sponge, respectively. This relationship indicates that the evaporation rate monotonically augments with the thickness of a sponge in a dark condition, which is due to the increase of total surface areas of a sponge with its thickness. Here, the side wall provides effective water transpiration paths that supports a proportional evaporation rate in a dark condition, which also accounts for the enhanced evaporation rate of thicker $MoS_2$/C @ PU sponges under solar illumination in **Figure 3b**.

With the above quantitative experiments, we can confirm and understand that solar steamers with 3D evaporation paths and enhanced surface areas provide much greater evaporation rates than those with 2D evaporation paths. The surface-area effect of the solar-steaming sponges is further demonstrated by sculpturing a $MoS_2$/C @ PU sponge into a spoke-like structure (10 mm in thickness, inset in **Figure 3f**). As shown in **Figure 3f**, owing to the enhanced evaporation surfaces along the side walls, the spoke-like structure provides evaporation rates to at least 1.84 kg m$^{-2}$ h$^{-1}$ in the indirect contact design [**Figure 2a-3**], and 1.95 kg m$^{-2}$ h$^{-1}$ in the non-contact design [**Figure 2a-5**] under 1 sun, which are 16.5% and 16.1% improvements compared to those rates offered by the cylindrical structures of the same thickness, respectively. The above results indicate that optimizing the geometrical factors is a highly viable approach to improve the performance of solar steamers.

2.2.3. High Solar-Thermal Energy Conversion Efficiency

Next, we determined the solar-thermal energy conversion efficiency of the $MoS_2$/C @ polyurethane Sponges ($\eta$) assembled in the five different setups in **Figure 2a**. The efficiency is given by $\eta = \dot{m}H_{Lv}/P_{in}$,[17, 29] where $\dot{m}$ is the mass flux generated by the absorber ($\dot{m} = \dot{m}_{\mathrm{w/absorber\ under\ sun}} - \dot{m}_{\mathrm{w/absorber\ in\ dark}}$), $P_{in}$ denotes the power density of solar irradiation on the absorber surface (1 kW $m^{-2}$ for 1 sun), and $H_{LV}$ is the enthalpy of liquid-vapor phase transition, including both the sensible heat and latent heat. Here we utilize the latent heat only to obtain a conservative result, detailed in Table S1. Accordingly, energy transfer efficiencies of $MoS_2$/C @ PU sponges with different thicknesses and water contact conditions are provided in **Figure 4a**. Unlike the evaporation rate, which monotonically increases with the thickness of a sponge, the solar-thermal energy conversion efficiency reaches a maximum value of 88% at an optimal thickness of 10 mm (setup 5, non-contact design), and levels off at 84% at a thickness of 16-19 mm.

It is known that the efficiency of solar-thermal energy conversion is governed by thermal loss of the solar steamer via several routes,[6, 48] including thermal radiation $\Phi \propto A \cdot (T_{ab}^4 - T_{am}^4)$, thermal convection to air $Q_{cv} \propto A \cdot (T_{ab} - T_{am})$, and thermal conduction to bulk water $Q_{cd} \propto V \cdot \Delta T$, where $T_{ab}$ and $A$ are the average operating temperature and surface area of the solar steamer, respectively, $T_{am}$ is the ambient temperature, $V$ is the volume of stored water, and $\Delta T$ is elevated temperature of bulk water. When increasing the sponge thickness from a very small value, solar radiation can essentially heat up the entire sponge along the thickness. The thermal loss is dominated by thermal irradiation ($\Phi$) and convection to air ($Q_{cv}$). The increase in thickness of the solar-steaming sponge provides lower average operating temperature ($T_{ab}$) as determined in **Figure 4b**, which reduces the thermal irradiation and convection. Therefore, the conversion efficiency augments with thickness of the sponges in the beginning. However, a further increase in thickness allows a larger bulk water storage, in which the solar irradiation cannot heat up the entire sponge uniformly. As shown in Figure S6, the cross-sectional IR imaging of 19 mm $MoS_2$/C @ PU exhibits a strong temperature drop along the direction of thickness compared to that of 10 mm composite [**Figure 2d-2**], which generates internal fluidic convection and thus energy loss, accounting for the reduced solar-thermal efficiency after reaching the optimal thickness. Consequently, the sculptured spoke-like structure (10 mm in thickness) could provide an ultrahigh energy efficiency of 91.5% in the indirect contact design, or 94% in the non-contact design under 1 sun [Table S1]. This is because at the optimal thickness of 10 mm, the spoke-like structure contains less bulk water than the cylindrical structure with similar diameter (25 mm), which results in much less energy loss and thereby a higher energy conversion efficiency.

Finally, serving as both a freestanding water reservoir and a solar steamer, the $MoS_2$/C @ PU sponge offers over 80% porosity for water storage and transportation. The time dependent evaporation dynamics for a series of fully-water-absorbed $MoS_2$/C @ PU sponges is tested continuously under 1 sun until all water evaporates. As observed in **Figure 4c**, mass loss in the first 30 min under solar illumination is lower than the following before a stable operation temperature can be established. Then the rate of mass change remains almost constant before sharply dropping to zero in the last 30 min when water stored in the sponge is completely gone. The 16 mm $MoS_2$/C PU sponge can unceasingly generate water vapor for ~8 hours at an almost constant rate, demonstrating the high reliability of our $MoS_2$/C PU solar steamers.  Note that the 8-hour operation time removes the necessity to replenish water during the day time. Moreover, the cumulative water mass loss linearly increases with sponge thickness. From the slope one can readily determine a water storage capacity of 869 kg $m^{-3}$ [**Figure 4d**], which is close to the density of pure water due to the high porosity of the sponge.

### 2.3. Applications in Water Treatment

The overall excellent performances of $MoS_2$/C @ PU sponges in solar-steaming make them promising candidates for water purification and treatment at homes and large facilities owing to the ease in fabrication and relatively low materials cost of ~\$0.93/$in^3$ [Table S2]. For demonstrations, we designed a solar steam collection system as shown in **Figure 5a** and Figure S6, respectively. In this setup, similar to setup 3 in Figure 2, contaminated water is stored in the inner vial covered by a thermally insulating layer wrapped with cellulose paper. A $MoS_2$/C @ PU sponge is stacked on top of the insulating layer. The produced steam condenses on the cone-shaped ceiling and flows along into the outer water basin connected with a faucet.

Water contaminated with bacteria as well as metal ions is tested. After one-step solar-driven steaming, as shown in **Figure 5b**, water taken from the Colorado River in Austin, Texas, is reduced in both alkalinity and hardness. The alkalinity drops from 180 ppm to 40 ppm and hardness is reduced from 200 ppm to 50 ppm. The test of bacteria changes from positive to negative, demonstrating the effective disinfection function of our solar steaming device **[Figure 5(b, c)]**. Furthermore, the solar steaming device shows a high efficiency in removing heavy metal ions of nickel. As shown in the inset in **Figure 5d**, the color of nickel ions solutions (0.5 M) go from green to clear after one-step solar steaming. Quantitative analysis determines that the concentration of nickel ions ($Ni^{2+}$) is reduced by ~ 99.6%, from 27 g $L^{-1}$ to 1 mg $L^{-1}$ [**Figure 5d**].

Mercury, which is known for its bio-accumulativeness and toxicity to central nerve systems, poses significant dangers to natural ecosystems and human being. The upper limit of Mercury in drinkable water is only 2 μg $L^{-1}$ (2 ppb) as set by the US EPA, much lower than the safety levels of most metal ions, *i.e.* 500 ppm for Na, 100 ppm for Ca, 10 ppm for Mg, and 1.3 ppm for Cu.[12, 26] However, no solar steamers have demonstrated the removal of Mercury, not to mention to the safe level set by the EPA in one step. In this work, leveraging the high binding affinity between $MoS_2$ and Mercury, the $MoS_2$/C @ PU sponge is utilized for efficient mercury removal with solar energy. Here, we chose mercury chloride of 200 ppb for test, which has the same level of mercury as that of wastewater from polyvinyl chloride production, one of the major sources of anthropogenic mercury.[37] The investigation starts with testing the mercury removal functions of suspending amorphous carbon (AC) and $MoS_2$/C microbeads as shown in **Figure 5e**. Indeed, by vigorous mixing and centrifuging, microbeads made of $MoS_2$/C (10 mg) can effectively reduce the mercury concentration from 200 ppb to less than 2 ppb (details in experimental section and schematic in Figure S7), while AC microbeads of the same amount do not remove mercury at all. It is known that the hydrothermally synthesized $MoS_2$ contains $H^+$ ($H_xMoS_2$) to balance its negative charges.[37] The removal of mercury by $MoS_2$/C could be attributed to the exchange of $Hg^{2+}$ with $H^+$ ions and formation of strong Hg-S bonding, following the reaction of $\mathrm{H}_x\mathrm{MoS}_2 + y\mathrm{HgCl}_2 \rightarrow \mathrm{Hg}_y\mathrm{H}_{(x-2y)}\mathrm{MoS}_2 + 2y\mathrm{H}^+ + 2y\mathrm{Cl}^-$.[37, 60, 61] This reaction suggests the feasibility of using the $MoS_2$/C functionalized PU sponges for mercury removal during solar steaming.

Indeed, solar steaming with the $MoS_2$/C @ PU sponge can effectively reduce mercury ions down to 1ppb, meeting the standard of drinkable water (**Figure 5f**). Here, the hydrophilicity of $MoS_2$/C PU provides strong capillary force to pull up the mercury-polluted water and generates intimate surface contact, during which mercury ions can be adsorbed by the surface of coated $MoS_2$/C, mostly before water transitions to steam. Here, the mesoporous PU sponges with interconnected 3-D structure and controlled thickness ensure sufficient reactions of $MoS_2$/C with mercury ions. To the best of our knowledge, this is the first demonstration of a solar steaming based mercury decontamination device. Different from that of the mercury removal with $MoS_2$/C microbeads by mixing and centrifuging in suspension, the solar steaming approach can achieve the results in one operation, without suspicion of mercury-containing residuals. To further determine the role of $MoS_2$ in mercury removal, we directly steamed mercury contaminated water with 1 sun, which does not show reduction of mercury ions with a concentration of 200 ppb. In a comparison experiment, we found 3D graphene foams (GF) can offer a moderate mercury removal efficiency, reducing mercury from 200 ppb to 50 ppb, while not meeting the EPA standard of 2 ppb in one operation [**Figure 5e**].

## 3. Conclusion

In summary, $MoS_2$/C @ PU sponges are strategically synthesized and assembled as 3-D freestanding solar steamers, which offer one of the best water evaporation rates of 1.95 kg $m^{-2}$ $h^{-1}$ and an efficient one-step mercury removal function. The high performance and unique functions can be attributed to multiple factors designed in the materials system. Owing to the broadband optical absorption of amorphous carbon beads, $MoS_2$/C @ PU sponge exhibits excellent solar absorption up to ~98% in a wide wavelength range from 200 to 2500 nm. The 3-D vapor evaporation path of the sponge, enabled by its adjustable aspect ratio, provides an optimized water evaporation rate as high as 1.95 kg $m^{-2}$ $h^{-1}$ under 1 sun. The low thermal conductivity of the sponge contributes to heat localization to the surface of $MoS_2$/C beads for dually-functional water steaming and synergistic mercury removal. The 3-D porous structure directly stores water in small porous compartments, which could effectively suppress conventional thermal losses, offering an energy efficiency of 88%. When applied to decontaminate natural river water, the $MoS_2$/C @ PU successfully eased alkalinity, reduced hardness, and removed bacteria. The suggested solar steaming system is particularly effective in removing mercury ions from contaminated water, offering an efficiency of ~99.6% in one operation, to a level of 1 ppb. The multifaceted innovation reported in this work, including material processing and assembling, water handling, and mercury removal, could inspire new paradigms of solar steaming technologies for applications relevant to water sterilization, detoxification, and desalination.

## 4. Experimental Section

*Synthesis of $MoS_2$/C microbeads*. A mixture of deionized (DI) water (75 mL), glucose (0.3 g), sodium molybdate ($Na_2MoO_4 \cdot 2H_2O$, 0.3 g) and thiourea (0.6 g) is added into a Teflon-lined stainless-steel autoclave and kept in an oven at 220℃ for 24 hours. After cooling to room temperature, the black precipitates of $MoS_2$/C are centrifuged, washed with DI water and ethanol alternately three times, and finally dried in a vacuum oven at 60℃. For control experiments, amorphous carbon microbeads are prepared via a similar route without the use of sodium molybdate.

*Fabrication of $MoS_2$/C @ PU*. A cylindrical PU sponge (25 mm in diameter, 10 mm in height) is successively soaked in poly(diallyldimethylammonium chloride) solution (PDDA, 4 wt%), and poly(sodium-p-styrenesulfonate) solution (PSS, 4 wt%) with ultrasonication for 10 minutes, rinsed with DI water, and dried in blowing air. This step is repeated twice and concluded with another layer of PDDA coating. Then the five-polyelectrolyte-layer coated PU sponge is immersed the $MoS_2$/C ink comprised of $MoS_2$/C (0.25 g) and ethanol (50 mL), followed by overnight shaking on a vortex mixer, and 6-hour drying in vacuum oven at 60°C.

*Mercury removal by suspending microbeads*. Amorphous carbon and $MoS_2$/C microbeads (10 mg) are respectively added to mercury chloride solutions (200 ppb, 10 mL) and mixed in a vortex shaker for 4 hours. The microbeads are then separated from the solution by centrifuging at 4000 rpm for 5 min.

*Characterization*. SEM characterizations are conducted with an FEI Quanta 650 ESEM and Hitachi S5500 SEM. Raman spectra are measured using Witec micro-Raman spectrometer alpha 300. XPS results are obtained by a Kratos X-ray Photoelectron Spectrometer. Rigaku MiniFlex 600 II characterized XRD spectra. The zeta potentials are tested by Dynamic Light Scattering Zetasizer Nano ZS. The optical absorption spectrum is measured by a Cary 5000 UV-VIS NIR spectrometer from 200 to 2500 nm. IR images are recorded by a Fluke Ti100 camera to observe the temperature distribution of $MoS_2$/C @ PU under 1 sun illumination. The solar simulator is an Oriel LCS-100 94011A and calibrated by Newport 91150V. The mass change is measured by an electronic balance (ADAM NBL 223e, 0.001g) in real time. The concentration of trace elements is analyzed by inductively coupled plasma mass spectrometry. The alkalinity, hardness and bacterial tests are carried out with water testing kits offered by Test Assured (USA).

**ASSOCIATED CONTENT**

**Supporting Information**

Supporting information is available from the Wiley Online Library or from the author.

**Corresponding Author**

*dfan@austin.utexas.edu

**ACKNOWLEDGMENTS**

The authors are grateful for the support of Welch Foundation (Grant No. F-1734) and the National Science Foundation (Grant No. CMMI 1563382).

**CONFLICT OF INTEREST**
The authors declare no conflict of interest.

**Figures**

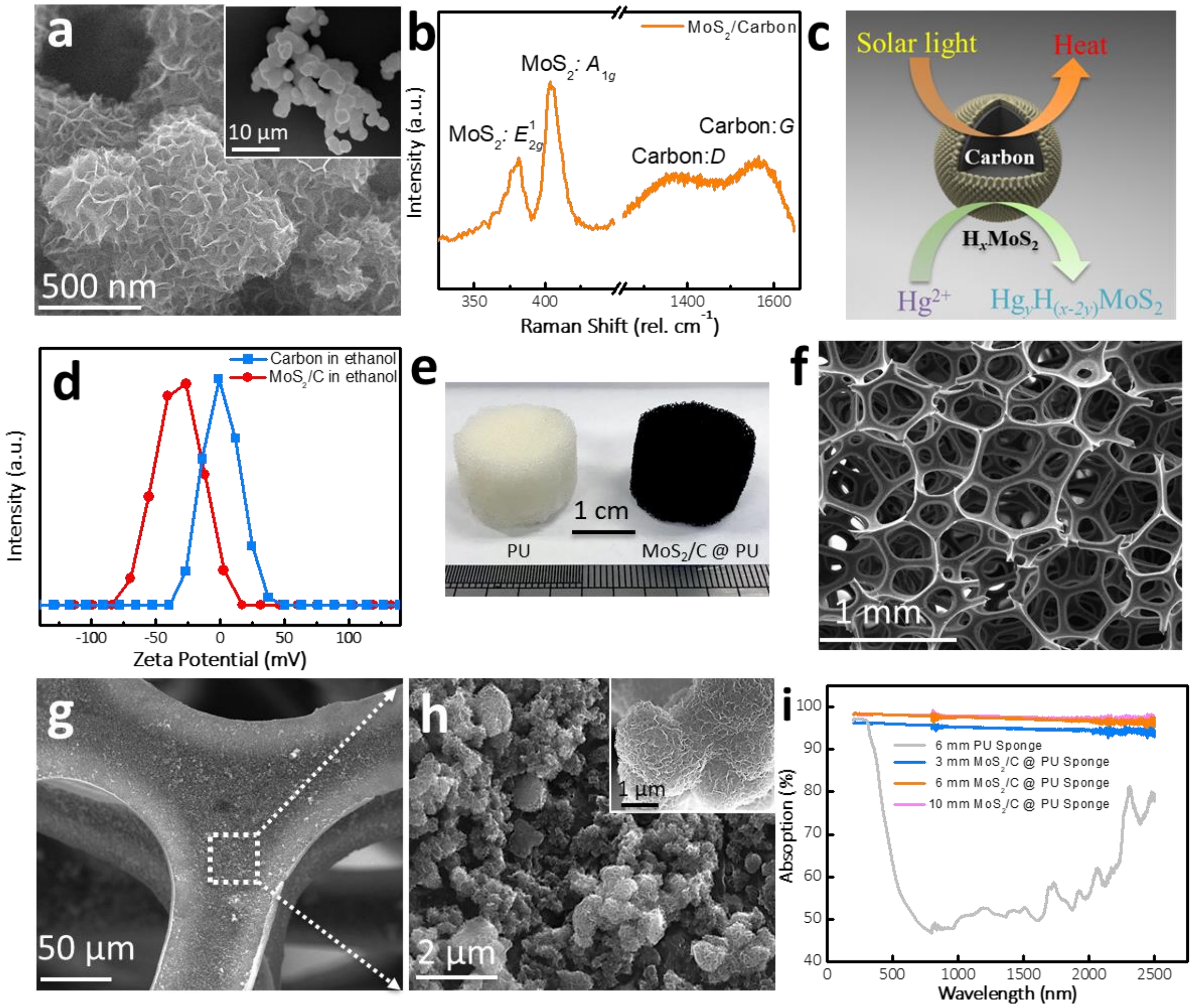

**Figure 1.** Characterizations of $MoS_2$/C @ PU sponge for solar steaming. (a) SEM and (b) Raman spectrum of $MoS_2$/C microbeads. (c) Schematic of structure and functions of $MoS_2$/C microbeads. (d) Zeta potentials of $MoS_2$/C (in red) and amorphous carbon (AC) microbeads (in blue) in ethanol. (e) Photographs of PU sponges before and after assembling $MoS_2$/C microbeads. (f-h) SEMs of $MoS_2$/C @ PU sponge. (i) Optical absorption spectra of PU and $MoS_2$/C @ PU sponges at different thicknesses.

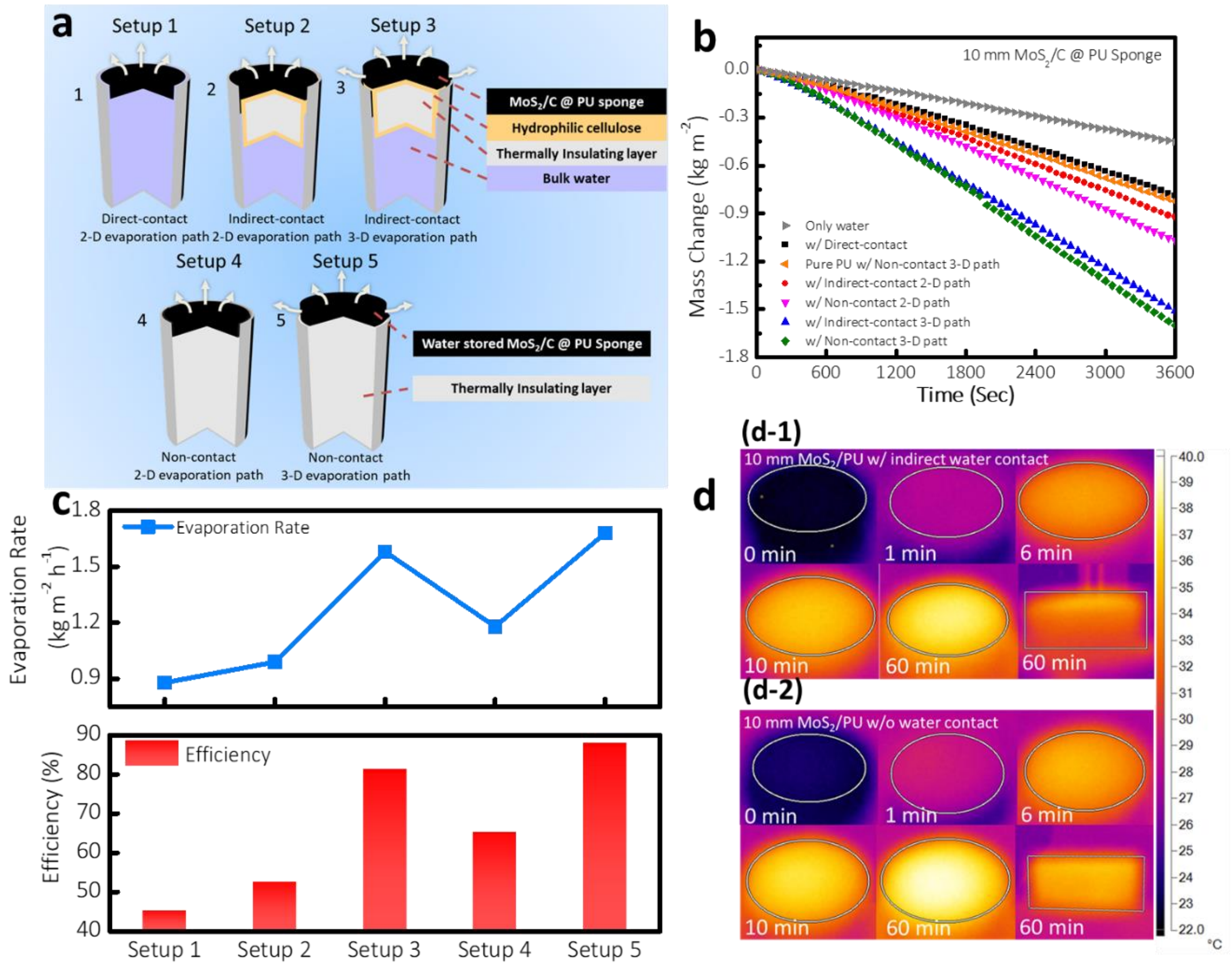

**Figure 2.** Characterization of solar steaming performance of $MoS_2$/C @ PU sponge (10 mm in thickness). (a) Schematic of different solar steam test designs: setup 1 to setup 5. (b) Cumulative mass change of water obtained with different test designs under 1 sun. (c) Solar-steaming evaporation rate and solar-thermal energy conversion efficiency obtained with different designs. (d) Selected infrared photos of a 10 mm $MoS_2$/C @ PU sponge (d-1) with indirect water contact (setup 3 in a) and (d-2) without water contact (setup 5 in a), the first five photos are top views and the sixth photo is a cross-sectional view.

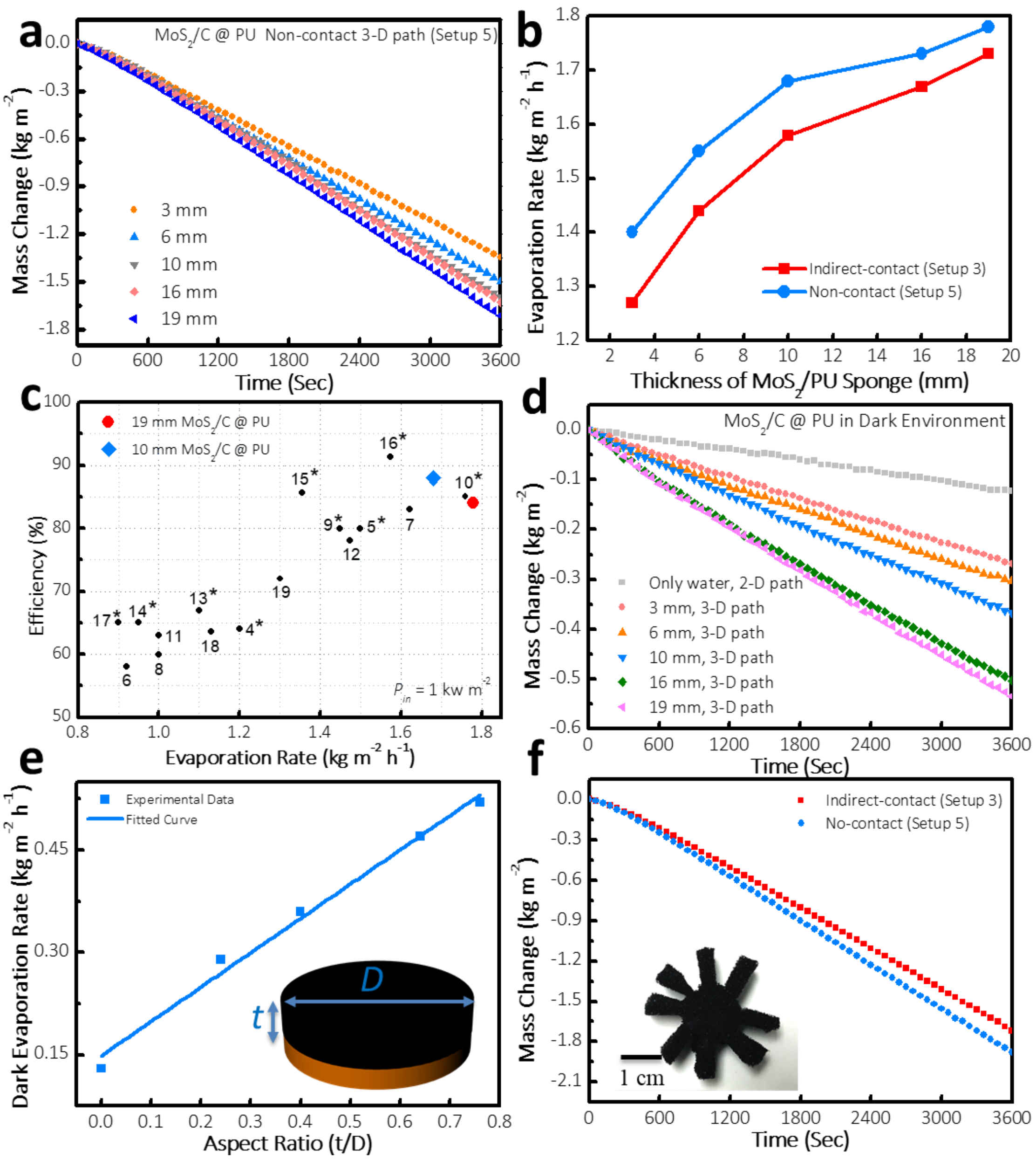

**Figure 3.** Characterization and analysis of the geometry effects of the 3-D solar steamers to their performances. (a) Cumulative mass change of water versus time obtained from $MoS_2$/C @ PU sponges of different thicknesses in the non-water-contact design, setup 5, under 1 sun. (b) Evaporation rates of $MoS_2$/C @ PU sponges versus thicknesses in the setup 5 (non-water-contact design, in blue) and setup 3 (indirect water contact, in red). (c) Performance comparison in solar-thermal conversion efficiencies and water evaporation rates under 1 sun irradiation. Annotation numbers correspond to the references in Table S1. The asterisks (*) mean both sensible heat and latent heat are considered in the efficiency calculation. (d) Cumulative mass change of water versus time obtained from $MoS_2$/C @ PU sponges of different thicknesses in a dark condition (Setup 5, non-water-contact design). (e) Evaporation rate versus aspect ratio of the composite sponge in a dark condition. (f)

Cumulative mass change of water versus time obtained from spoke structured composite sponge under 1 sun, inset: photograph of the spoke structured sponge.

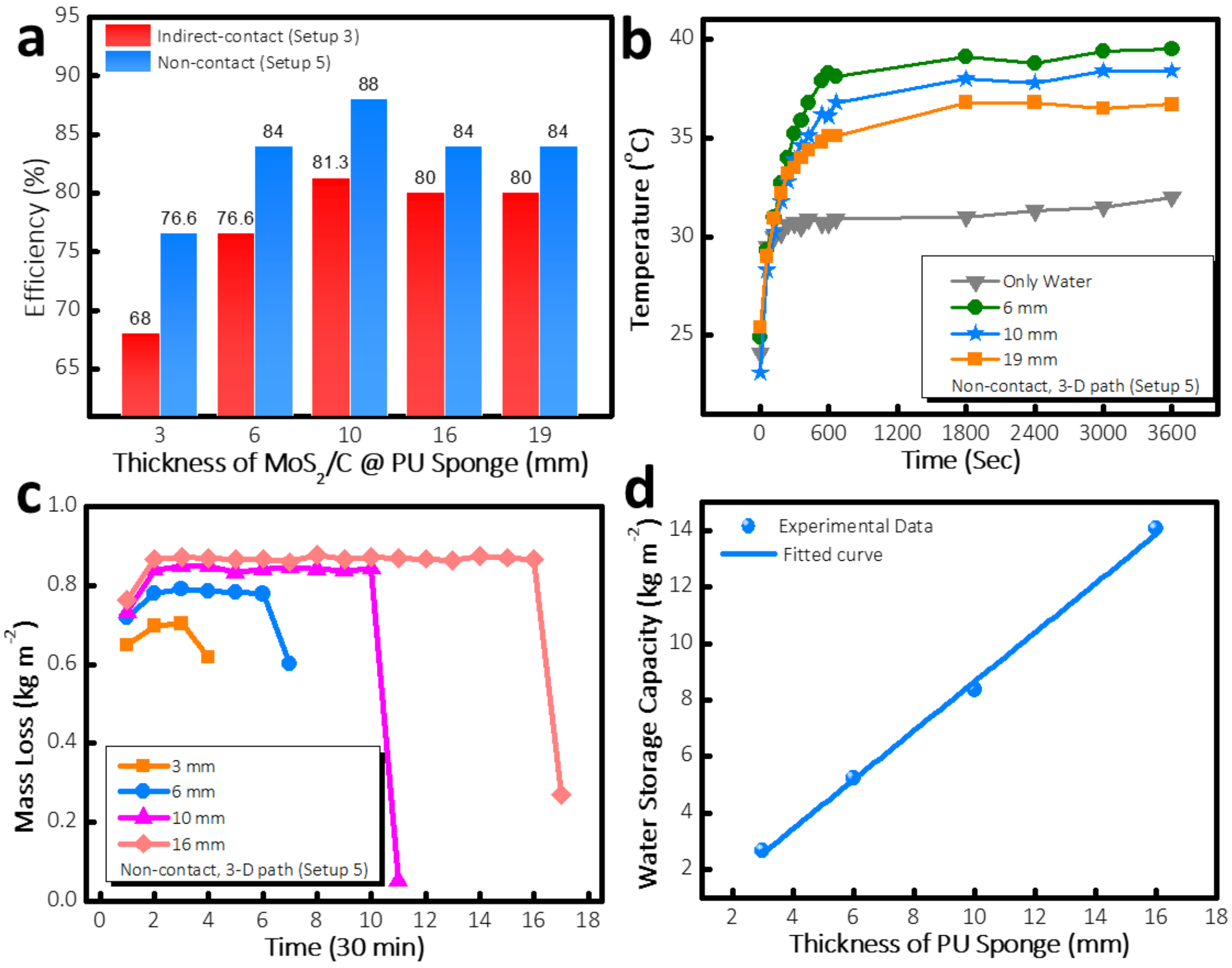

**Figure 4.** Solar steaming performances of 3-D $MoS_2$/C @ PU sponges of different thicknesses. (a) Solar steaming efficiency comparison among sponges of different thicknesses, configured in setups 3 and 5. (b) Maximum temperatures of the top surface of the composite sponge versus time. (c) Non-contact design: mass loss taken every 30 min from the $MoS_2$/C @ PU sponges with fully absorbed water under 1 sun. (d) Water storage capacity of $MoS_2$/C @ PU sponges versus thickness.

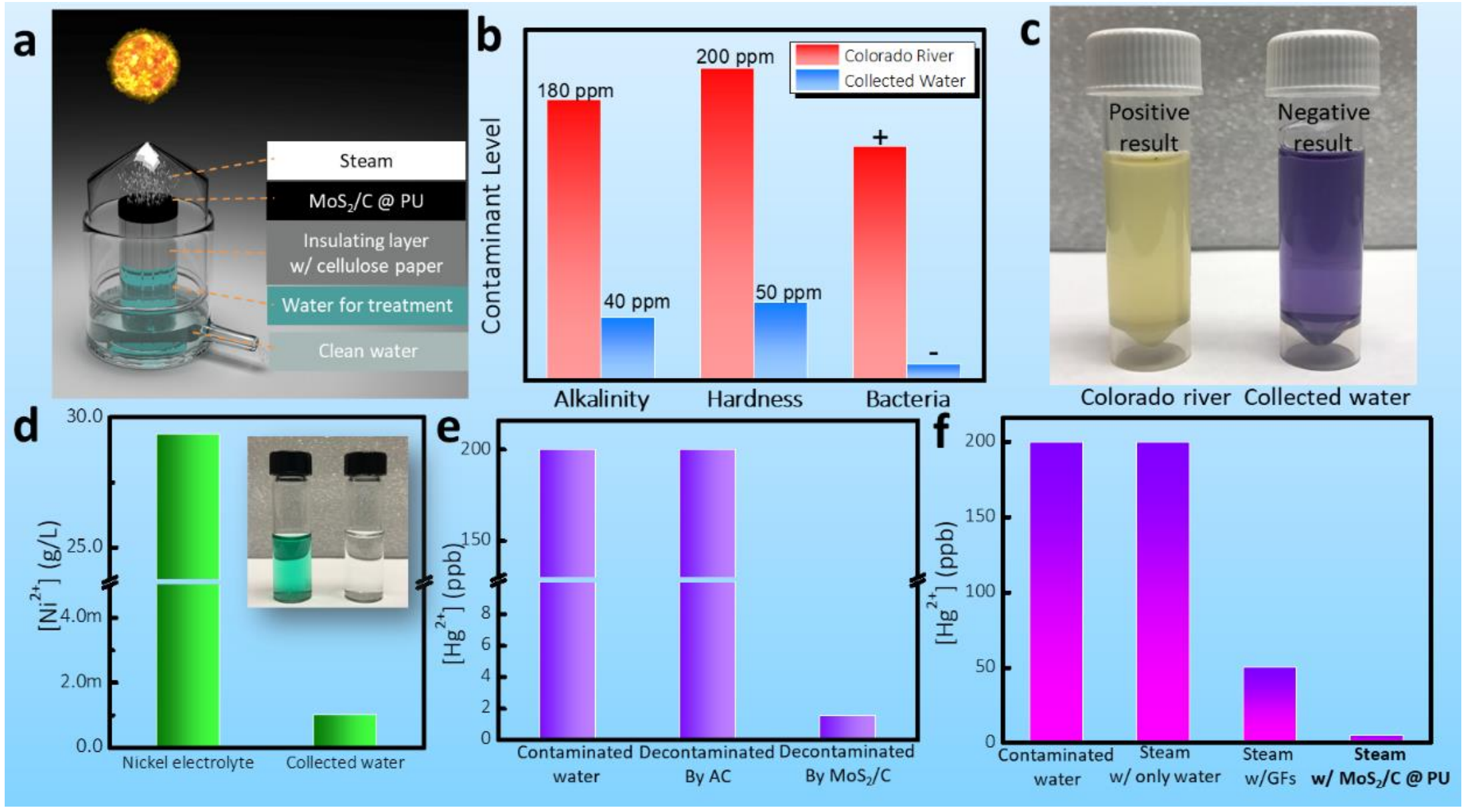

**Figure 5.** Applications of $MoS_2$/C @ PU solar steamers in a designed solar-steaming-collection system. (a) Schematic of the design of the solar steaming collection system. (b) Test of water from the Colorado River before and after solar steaming. (c) Bacterial tests of water from the Colorado River before and after solar steaming. Yellow color: bacterial positive; purple color: bacterial negative. (d) Concentration of Nickel ions before and after purification by solar steaming. Inset: original nickel containing solution (left) and water after steaming (right). (e) Comparison of Mercury removal functions by suspended $MoS_2$/C microbeads and amorphous carbon (AC) microbeads. (f) Comparison of effectiveness of Mercury removal by simple solar steaming, solar steaming assisted with graphene foams, and with $MoS_2$/C @ PU sponges.

# Supporting Information

**$MoS_2$/C @ Polyurethane Composite Sponges for Synergistic High-Rate Solar Steaming and Mercury Removal**

*Weigu Li,*[1] *Marshall C Tekell,*[2] *Yun Huang,*[3] *Karina Bertelsmann,*[1] *Max Lau,*[1] *and Donglei Fan*[1,3,*]

[1]Department of Mechanical Engineering, The University of Texas at Austin, Austin, TX 78712, USA

[2]Department of Chemical Engineering, The University of Texas at Austin, Austin, TX 78712, USA

[3]Materials Science and Engineering Program, Texas Materials Institute, The University of Texas at Austin, Austin, TX 78712, USA

## 1. Supporting Experimental Details

*Synthesis of 3D graphene foams for mercury removal control experiment.* The reaction starts with the annealing of commercial nickel foams (MTI Corporation, USA) in $H_2$ gas flow (20 sccm) at 700˚C for 40 minutes for the removal of surface oxides. Then ethylene ($C_2H_4$, 10 sccm) is introduced to grow ultrathin graphite on the nickel foam catalysts with a total pressure of 400 mTorr for controlled growth of graphite for 15 hours. Next, the temperature of the sample is reduced to room temperature in the original growth gas mixture. By selective etching Ni catalysts in a mixture of iron chloride ($FeCl_3$, 1 M) and hydrochloric acid (HCl, 2 M) at 60˚C overnight, freestanding graphite foams can be obtained. Next the ultrathin graphite foam is rinsed with deionized water and isopropanol a few times, and dried at 60˚C for 4 hr.

## 2. Supporting Figures

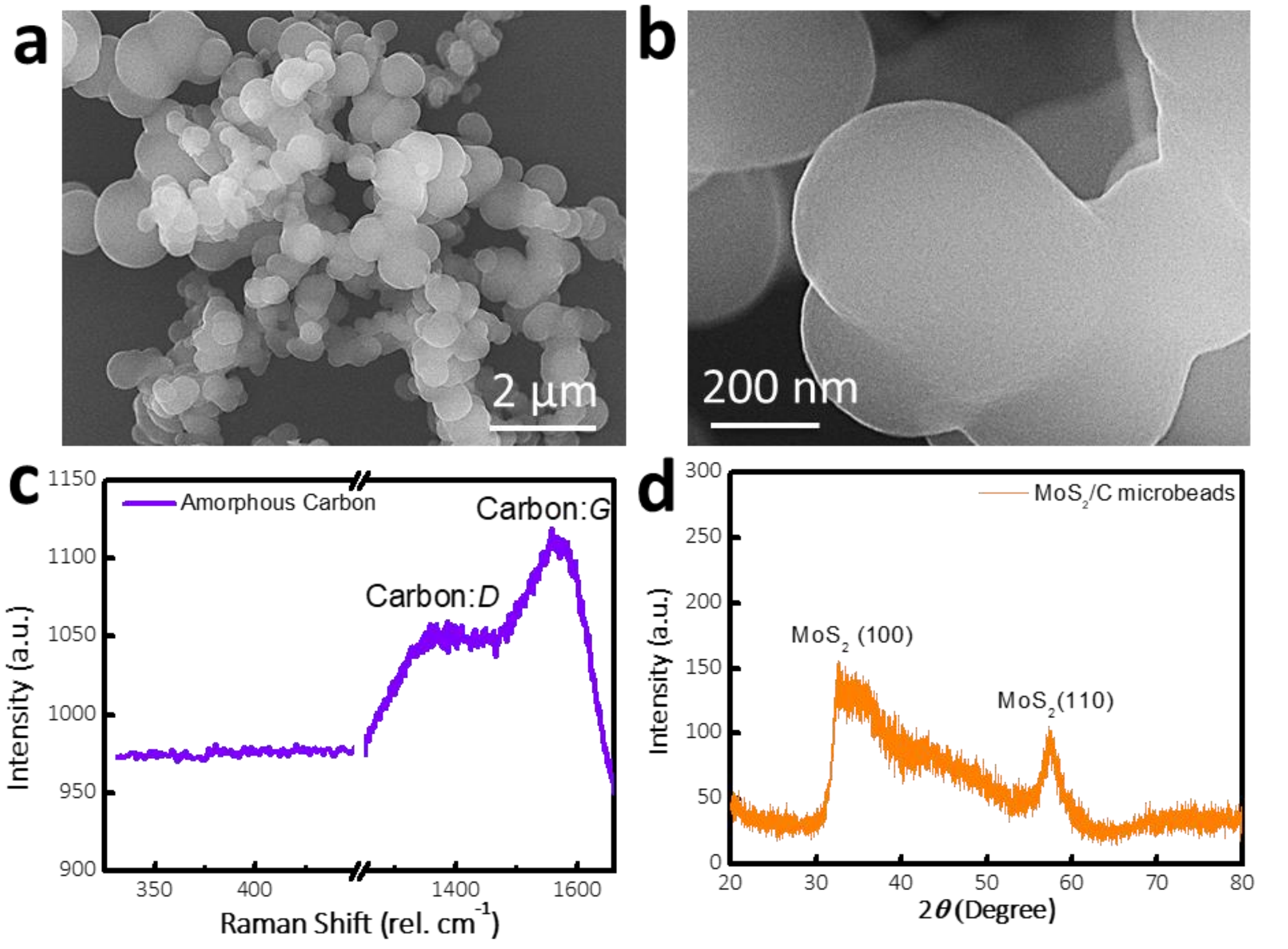

**Figure S1.** Characterizations of amorphous carbon microbeads. SEM (a, b) and Raman spectrum (c) of amorphous carbon. (d) XRD spectrum of $MoS_2$/C microbeads.

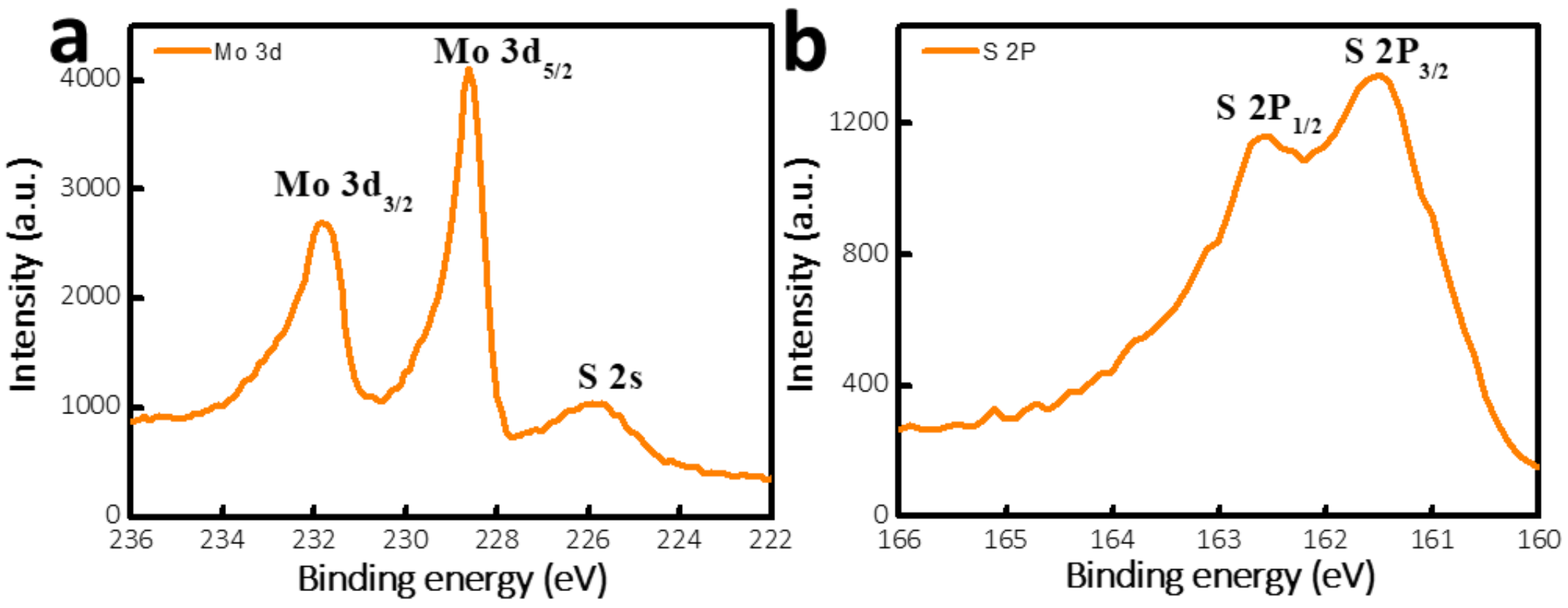

**Figure S2.** X-ray photoelectron spectroscopy (XPS) of the as-synthesized $MoS_2/C$ microbeads.

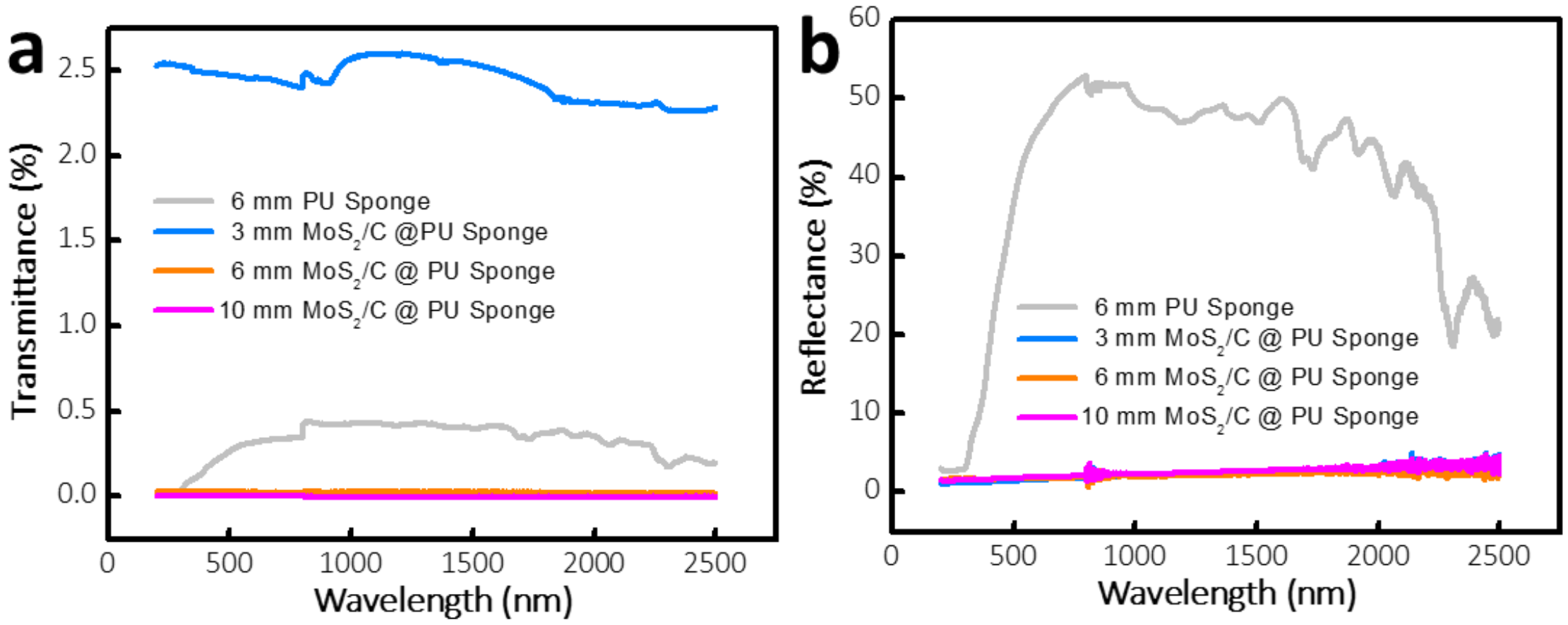

**Figure S3.** (a) Transmittance and (b) reflectance of PU and $MoS_2/C$ @ PU sponges of different thickness.

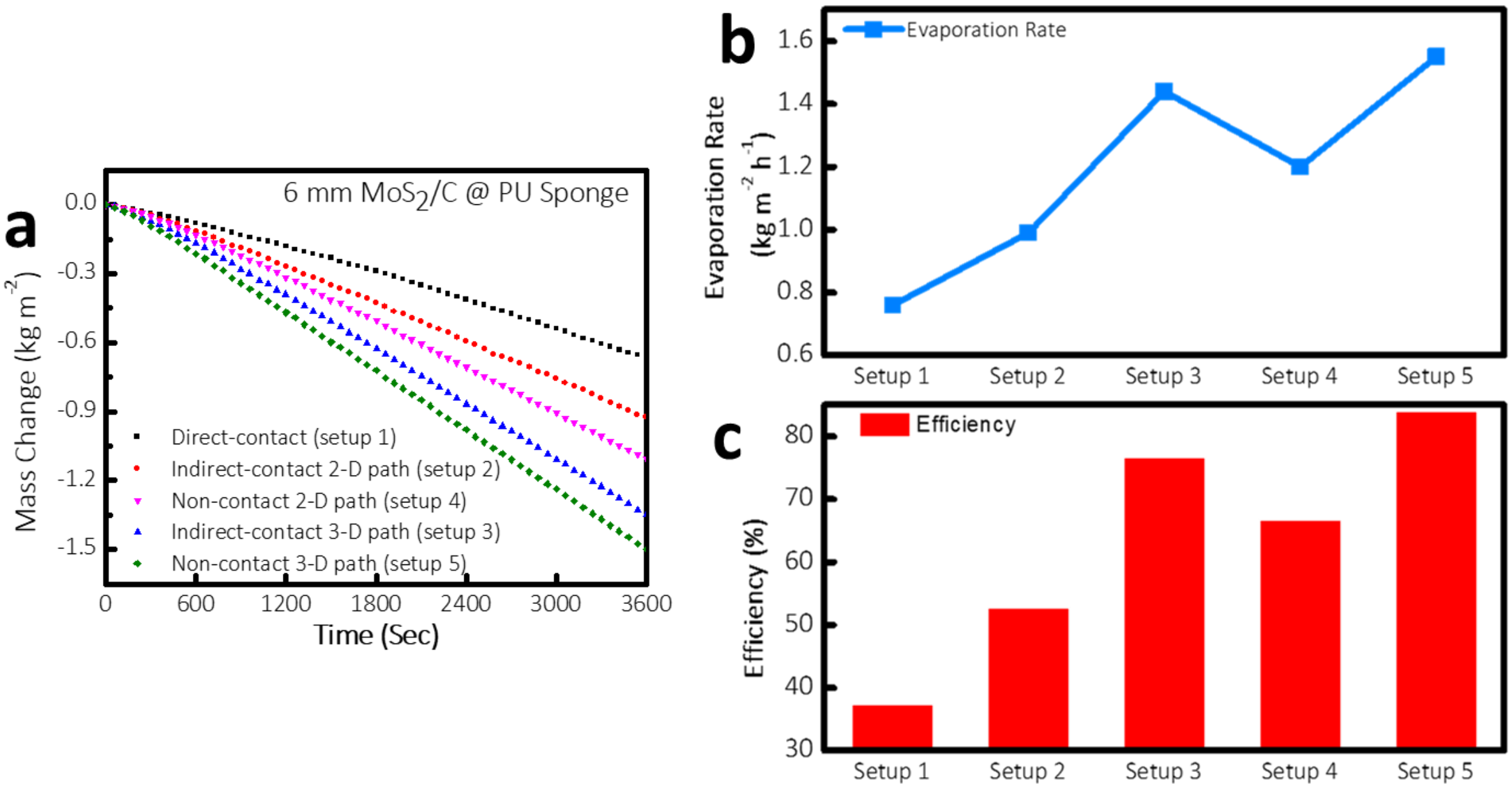

**Figure S4.** Characterization of solar steaming efficiency and evaporation rate of 6 mm $MoS_2/C$ @ PU sponge. (a) Accumulative mass change of water versus time with different

test designs under 1 sun. (b) Evaporation rate with different test designs. (c) Solar steaming efficiency with different test designs.

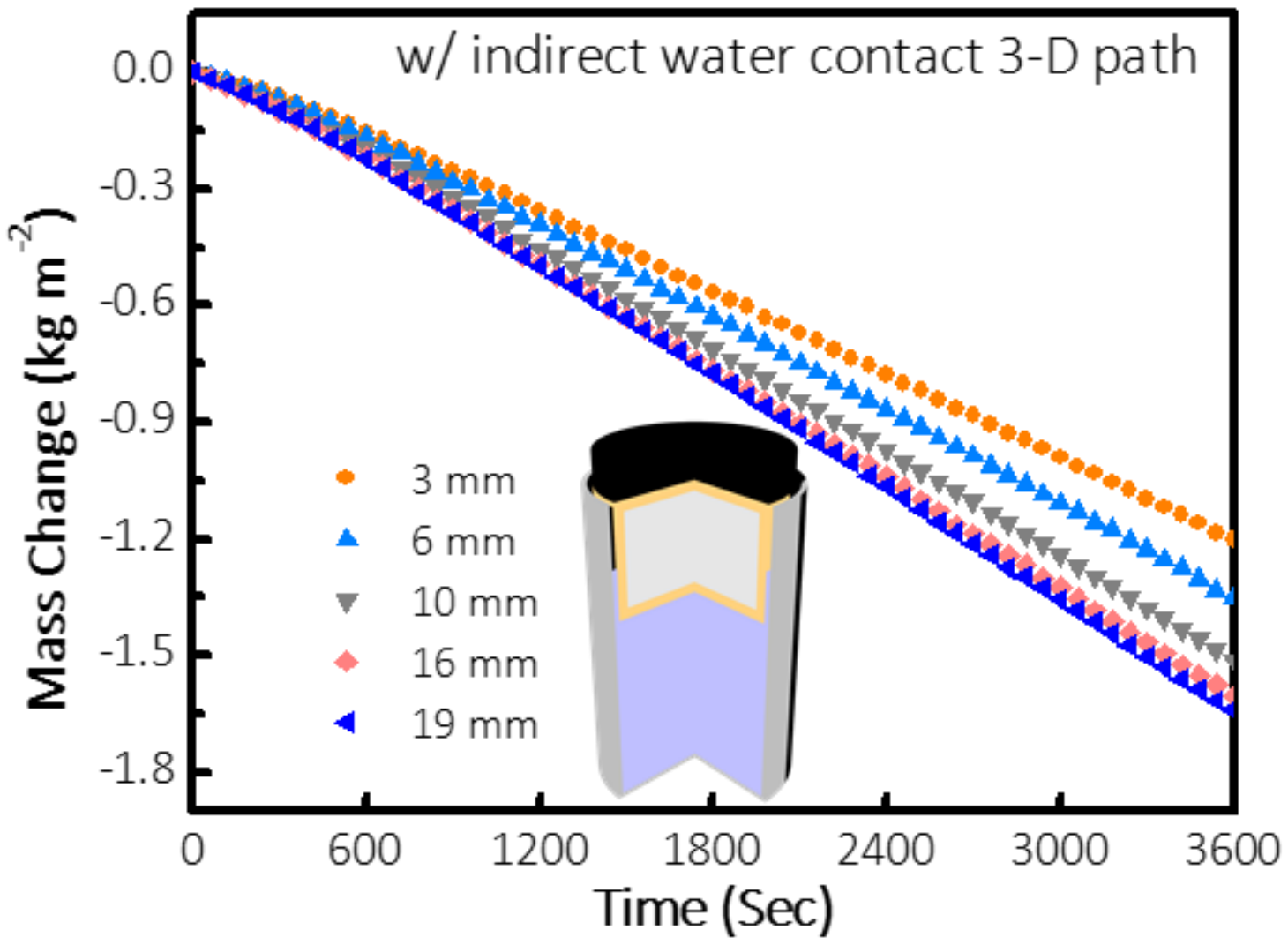

**Figure S5.** Accumulative mass change of water evaporated through $MoS_2$/C @ PU sponge steamers of different thicknesses with the design of setup 3 (indirect water contact and 3D evaporation) under 1 sun.

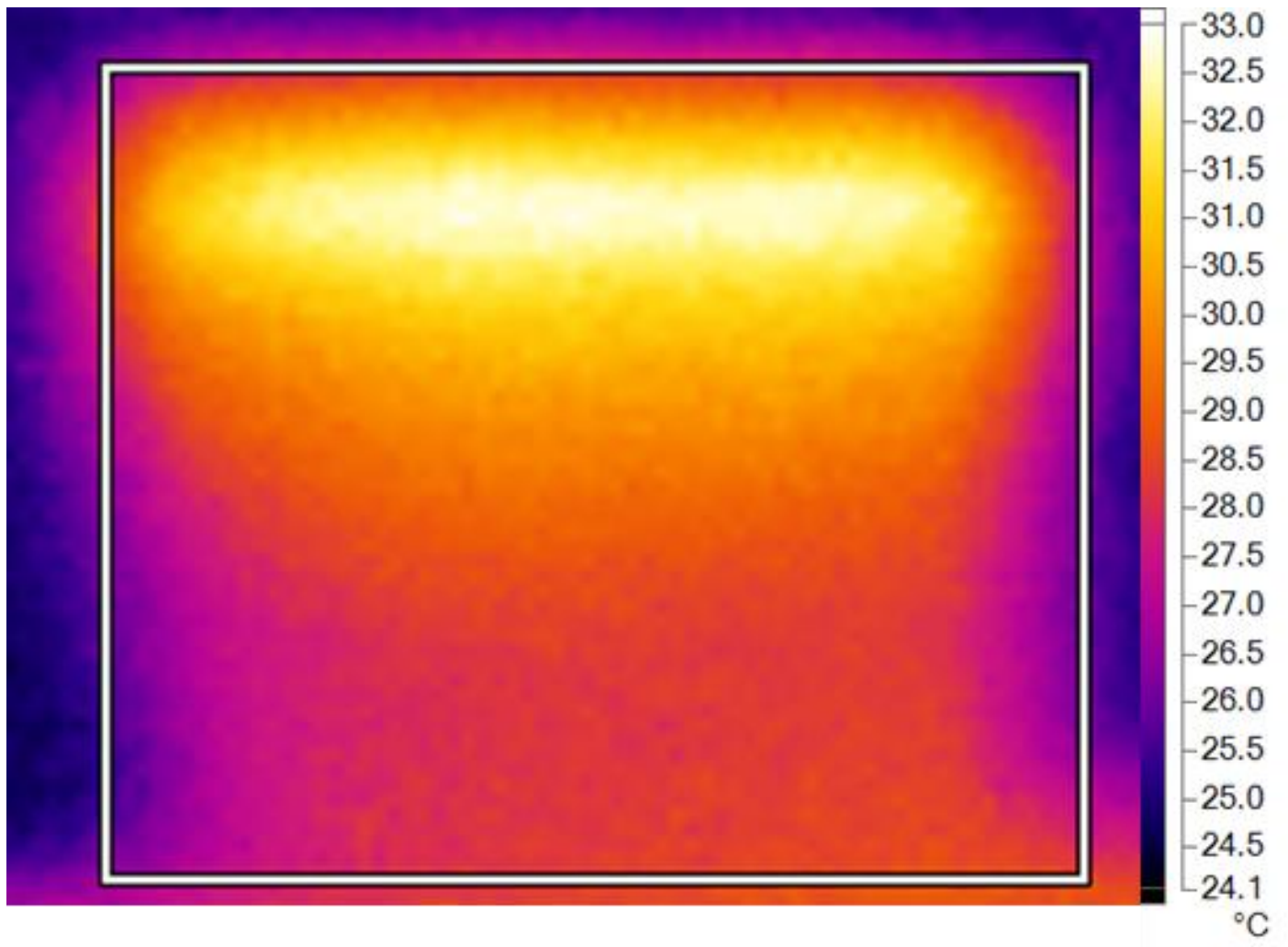

**Figure S6.** Cross-sectional infrared photograph of 19 mm $MoS_2$/C @ PU sponge without water contact (setup 5) after solar irradiation for 60 min.

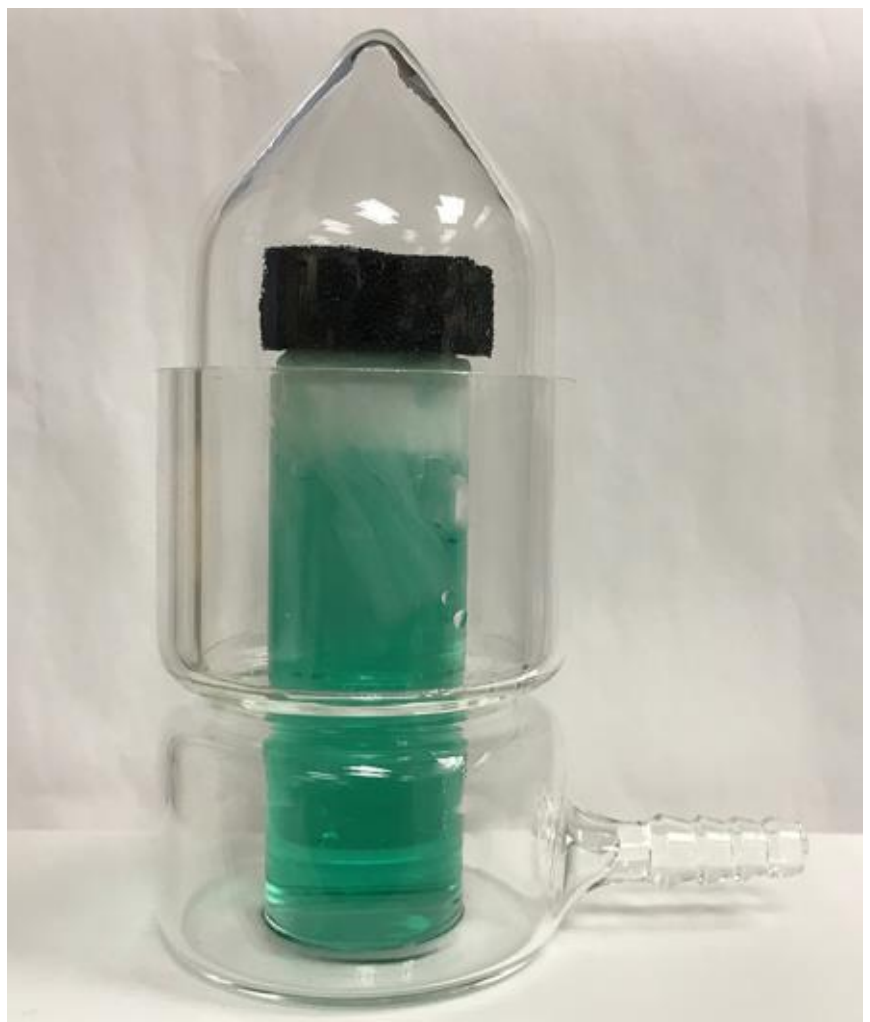

**Figure S7.** Photograph of a steam collection system for tests of water quality and ion removal after steaming with the $MoS_2$/C @ PU sponge in Setup 3 (indirect water contact and 3D evaporation) under 1 sun.

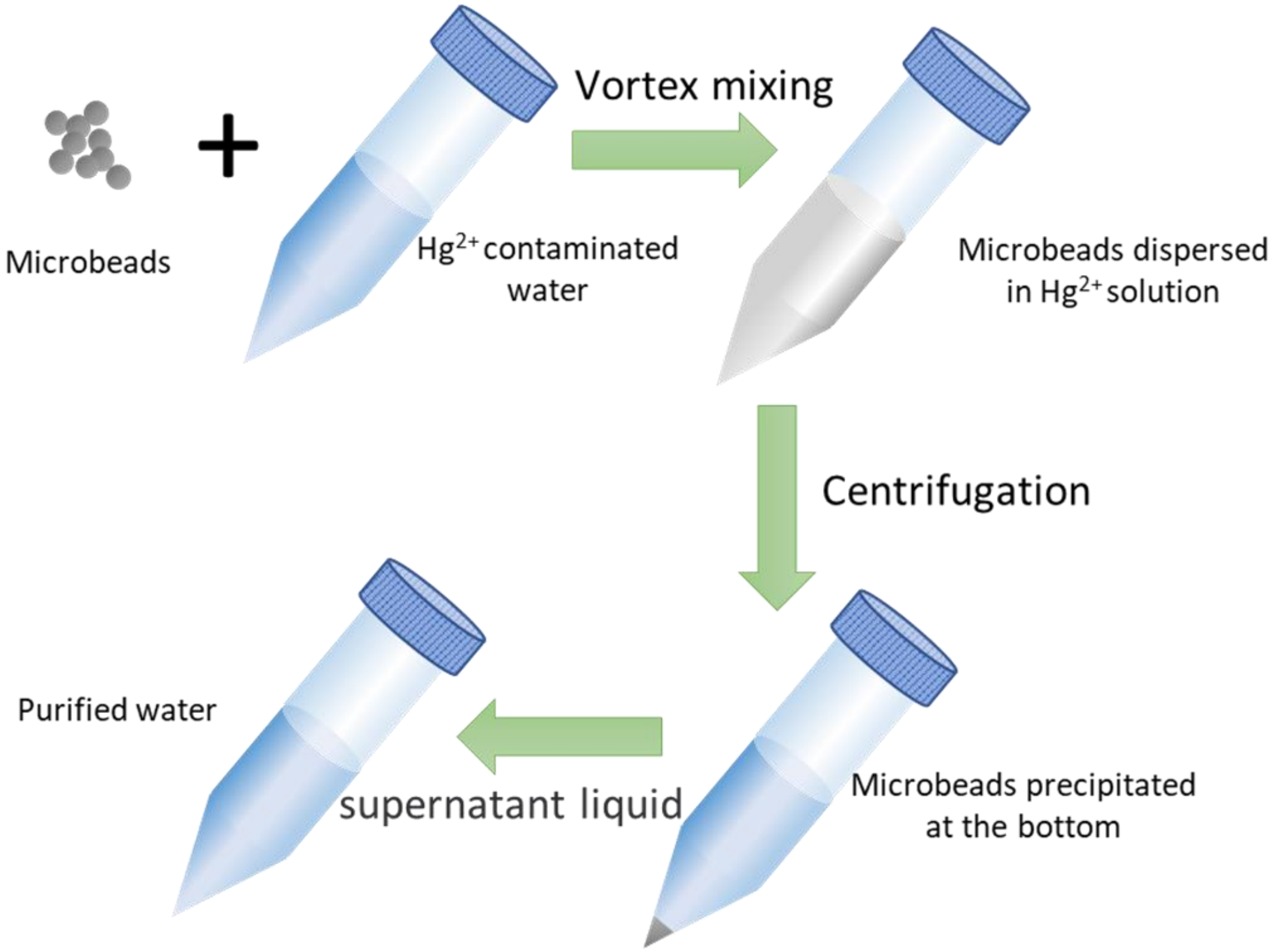

**Figure S8.** Control experiments: schematic of mercury removal process by mixing and removal of AC microbeads or $MoS_2$/C microbeads.

**Table S1.** Comparison of Solar Steaming Performances under 1 Sun Irradiation.

| Materials | Evaporation rate ($kg{\cdot}m^{-2}{\cdot}h^{-1}$) | Efficiency (%) | $H_{LV}$* | Ref. |
|---|---|---|---|---|
| Black gold membranes | 0.47 | 26 | S+L | Bae et al.[1] |
| Cauliflower-shaped hierarchical copper nanostructures | | 62 | L | Fan et al.[2] |
| Hollow carbon beads | 1.28 | | | Zeng et al.[3] |
| Exfoliated graphite and carbon foam | 1.2 | 64 | S+L | Ghasemi et al.[4] |
| Nitrogen-doped porous graphene | 1.5 | 80 | S+L | Ito et al.[5] |
| Polypyrrole coating on stainless steel | 0.92 | 58 | L | Zhang et al. [6] |
| Tailoring graphene oxide (GO)-based aerogel | 1.622 | 83 | L | Hu et al.[7] |
| Aluminum nanoparticles coated anodic aluminum oxide membrane | 1.0 | 60 | L | Zhou et al.[8] |
| 2-D GO film | 1.45 | 80 | S+L | Li et al.[9] |
| 3-D GO film | 1.76 | 85 | S+L | Li et al.[10] |
| Gold nanoparticles coated anodic aluminum oxide membrane | 1.0 | 63 | L | Zhou et al.[11] |
| Carbonized mushrooms | 1.475 | 78 | L | Xu et al.[12] |
| Plasmonic wood | 1.1 | 67 | S+L | Zhu et al.[13] |
| Carbon nanotubes (CNTs)/wood membrane | 0.95 | 65 | S+L | Chen et al.[14] |
| 3-D printed CNTs/GO/nanofibrillated cellulose composite | 1.355 | 85.6 | S+L | Li et al.[15] |
| Hierarchical graphene foam | 1.575 | 91.4 | S+L | Ren et al.[16] |
| Reduced GO/Polyurethane (PU) composite | 0.9 | 65 | S+L | Wang et al.[17] |
| Microstructured copper phosphate | 1.13 | 63.6 | L | Hua et al.[18] |
| Carbon black /polymethylmethacrylate/ polyacrylonitrile membrane | 1.3 | 72 | L | Xu et al.[19] |
| 3-D Carbon-coated paper | 2.20 | 100* | L | Song et al.[20] |
| Polyvinyl alcohol/polypyrrole gel | 3.2 | 94** | L | Zhao et al.[21] |
| **$MoS_2$/C @ PU sponge w/ indirect-contact** | **1.72** (19 mm PU) | 80.0 | L | **This work** |
| | | 81.8 | S+L | |
| | **1.58** (10 mm PU) | 81.3 | L | |
| | | 83.1 | S+L | |
| **$MoS_2$/C @ PU sponge w/ non-contact** | **1.78** (19 mm PU) | 84.0 | L | |
| | | 85.9 | S+L | |
| | **1.68** (10 mm PU) | 88.0 | L | |
| | | 89.9 | S+L | |
| **$MoS_2$/C @ PU spoke-like structure** **w/ indirect-contact** | **1.84** (10 mm PU) | **91.5** | L (conservative way in calculation) | |
| **w/non-contact** | **1.95** (10 mm PU) | **94.0** | L | |

| | | | | | |
|---|---|---|---|---|---|
| | | | | (conservative way in calculation) | |

* $H_{LV}$ is the liquid-vapor phase change enthalpy used for efficiency ($\eta = \dot{m}H_{LV}/P_{in}$). As shown in the Scheme S1, the evaporation of water undergoes two steps: (1) heating water from room temperature to the device operating temperature (sensible heat, S); (2) phase change from liquid to vapor at the operating temperaure (latent heat, L). However, different reports use different values.

** This work did not subtract the dark mass flux when calculating efficiency.

*** This work claims the water stored in the suggested gel structure evaporate as water clusters such that the vaporization enthalpy is much lower than conventional latent heat.

Calculation of Enthalpy by constructing an equivalent reaction route based on the Hess's Law in Thermodynamics:

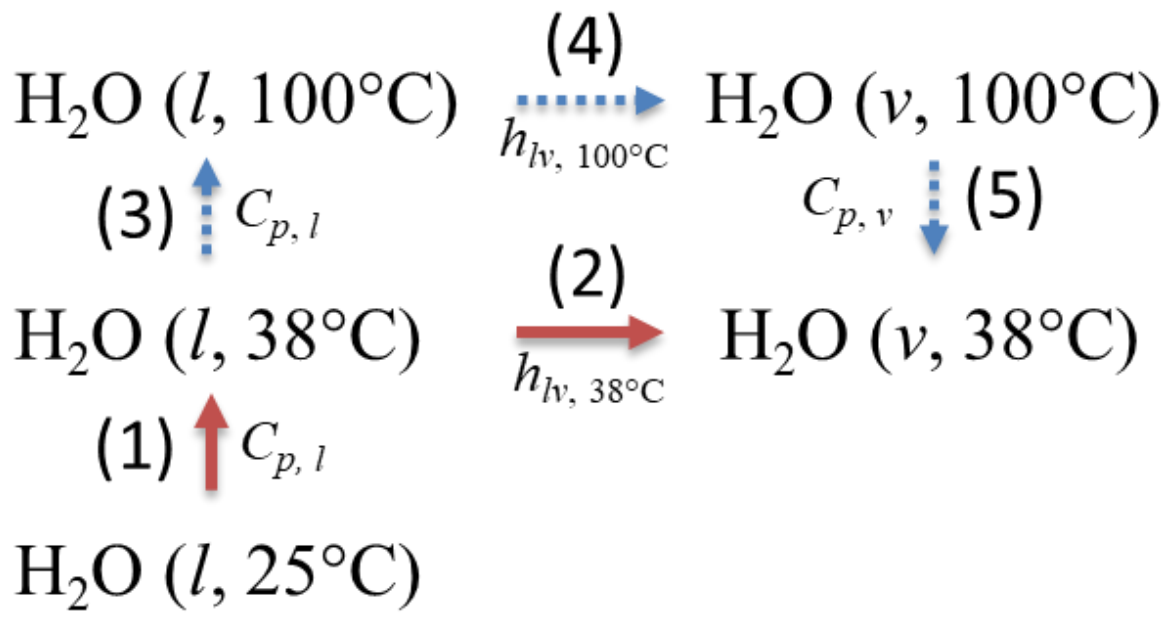

Scheme S1. Water evaporation routes.

In our work, we utilize the slope of the mass loss curve at the second 30 minute interval to estimate the evaporation rate, during which the water has aleady been heated to the operating temperaure. So we only consider the latent heat for a conservative efficiency caculation. In order to obtain an accurate latent heat at the specific temperature, we construct a multistep equilibrium reaction route according to the Hess' Law [Scheme S1].

$$h_{lv,38°\mathrm{C}} = \int_{38°\mathrm{C}}^{100°\mathrm{C}} C_{p,l}\,\mathrm{d}T \; + \; h_{lv,100°\mathrm{C}} + \int_{100°\mathrm{C}}^{38°\mathrm{C}} C_{p,v}\,\mathrm{d}T$$

Here, $h_{lv}$ is the latent heat, $C_{p,l}$ and $C_{p,v}$ are the heat capacitity of liquid water and water vapor, respecitivey. From the reference,[22] $h_{lv,100°C}$=2257 J·K$^{-1}$·g$^{-1}$, $C_{p,l}$=4.1813 J·K$^{-1}$·g$^{-1}$, $C_{p,v} = (3.470\; + 1.45\times10^{-3} \times T \; + \; 0.121 \times 10^{5} \times T^{-2}) \cdot R$ (J·K$^{-1}$·mol$^{-1}$), R =8.314 J·K$^{-1}$·mol$^{-1}$, T is the temperature in Kelvin scale. Therefore $\boldsymbol{h_{lv,38°C}}$**=2399.81 J·K$^{-1}$·g$^{-1}$**, which is used for the efficiency calculation in our work.

.

**Table S2.** Material cost estimation for 1 $MoS_2$/C @ PU composite.

| Chemicals | Vendor | Price | | Materials/ device | Price/ device |
|---|---|---|---|---|---|
| Glucose | Fischer Scientific | 500 g | $18.73 | 0.15 g | $0.01 |
| Sodium Molybdate | Acros Organics | 500 g | $130.25 | 0.15 g | $0.04 |
| Thiourea | Acros Organics | 1 kg | $196.82 | 0.3 g | $0.06 |
| Poly(sodium-p-styrenesulfonate) | Acros Organics | 500 g | $117.89 | 2 g | $0.47 |
| Poly(diallyldimethylammonium chloride) solution | Sigma Aldrich | 4 L | $130.00 | 10 mL | $0.33 |
| Polyurethane sponge | Sailrite | 4" · 36" · 82" | $147.95 | 1" · 1" ·1" | $0.02 |
| **Cost of $MoS_2$/C @ PU** | | | | | **$0.93** |